\documentclass{vgtc}                          

\graphicspath{{figures/}{pictures/}{images/}{./}} 

\usepackage{times}                     

\usepackage{tabu}                      
\usepackage{booktabs}                  
\usepackage{lipsum}                    
\usepackage{mwe}                       

\usepackage{mathptmx}                  
\usepackage[numbers,sort&compress]{natbib}
\usepackage{enumitem}
\usepackage{xspace}
\usepackage{float}
\usepackage{stfloats}

\onlineid{0}

\vgtccategory{Research}

\vgtcinsertpkg

\title{A Toolkit for Virtual Reality Data Collection}

\author{Tim Rolff\thanks{e-mail: tim.rolff@uni-hamburg.de}\\
        \scriptsize{University of Hamburg}\\
        \scriptsize{Department of Informatics}\\
\and Niklas Hypki\thanks{e-mail: niklas.hypki@uni-muenster.de}\\
        \scriptsize{University of Münster}\\
        \scriptsize{Department of Psychology}\\
\and Markus Lappe\thanks{e-mail: mlappe@uni-muenster.de}\\
       \scriptsize{University of Münster}\\
        \scriptsize{Department of Psychology}\\
\and Frank Steinicke\thanks{e-mail: frank.steinicke@uni-hamburg.de}\\
        \scriptsize{University of Hamburg}\\
        \scriptsize{Department of Informatics}\\
}

\teaser{
  \centering
  \includegraphics[width=0.94\linewidth]{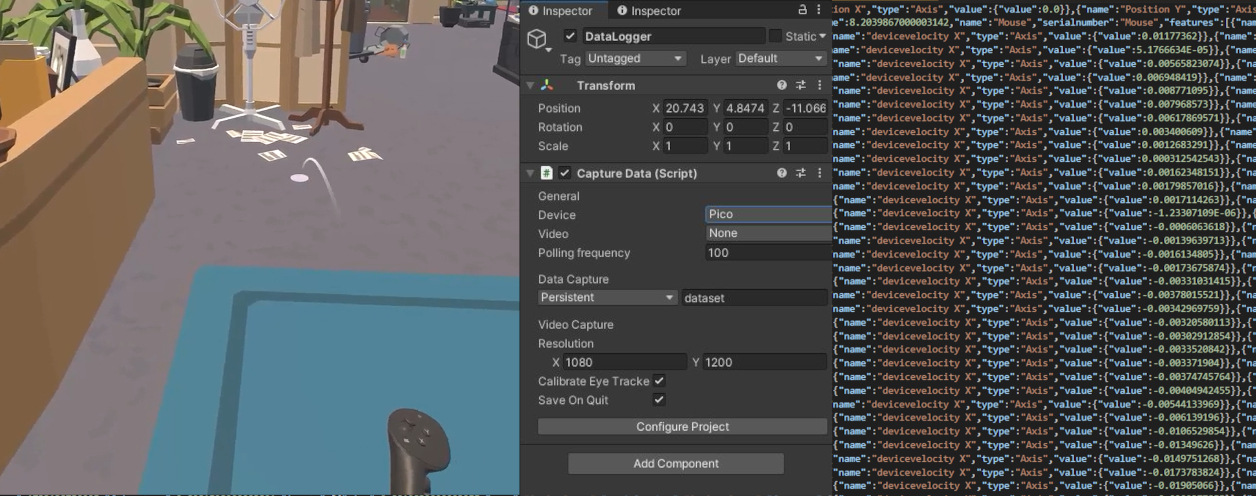}
  \caption{Illustration from an exemplary study setup showing (left) the virtual environment, (middle) our data collection tool, and (right) a snipped from the generated output.}
  \label{fig:teaser}
}

\abstract{
Due to the still relatively low number of users, acquiring large-scale and multi-dimensional virtual reality datasets remains a significant challenge. Consequently, VR datasets comparable in size to state-of-the-art collections in natural language processing or computer vision are rare or absent. However, the availability of such datasets could unlock groundbreaking advancements in deep-learning, psychological modeling, and data analysis in the context of VR. %
In this paper, we present a versatile data collection toolkit designed to facilitate the capturing of extensive VR datasets. Our toolkit seamlessly integrates with any device, either directly via OpenXR or through the use of a virtual device. Additionally, we introduce a robust data collection pipeline that emphasizes ethical practices (e.g., ensuring data protection and regulation) and ensures a standardized, reproducible methodology.
} 


\newcommand{\toolkit}{OXDR\xspace}
\newcommand{\fulltoolkit}{OpenXR Data Recorder\xspace}

\usepackage{flushend}
\usepackage{pdfpages}
\usepackage{appendix}
\usepackage{fancyhdr}
\usepackage{pdfpages}
\usepackage{xparse}
\usepackage{tikz}

\usetikzlibrary{automata, positioning, chains, shapes.symbols, fit, matrix}

\definecolor{uhhblue}{rgb}{0.08,0.443,0.733}
\definecolor{uhhgreen}{rgb}{0.08,0.733,0.443}

\makeatletter
\NewDocumentCommand\headerspdf{ O {pages=-} m }{
  \includepdf[%
    #1,
    pagecommand={\thispagestyle{fancy}},
    scale=.9,
    ]{#2}}
\NewDocumentCommand\secpdf{somO{1}m}{
  \clearpage
  \thispagestyle{fancy}%
  \includepdf[%
    pages=#4,
    pagecommand={%
      \IfBooleanTF{#1}{%
        \section*{#3}}{%
        \IfNoValueTF{#2}{%
          \section{#3}}{%
          \section[#2]{#3}}}},
    scale=.9,
    ]%
    {#5}}
\makeatother

\begin{document}


\firstsection{Introduction}

\maketitle

The vast availability of (open) internet data and media has become a cornerstone of modern machine learning, particularly fueling advancements in large language models (LLMs) and multimodal large language models (MLLMs). However, when applying machine learning to virtual reality (VR), data availability emerges as a critical limitation. Compared to the extensive and often terabytes-in-size datasets driving breakthroughs in fields like natural language processing (NLP) \cite{raffel2020exploring,patel2020introduction,weber2024redpajama} and computer vision (CV) \cite{russakovsky2015imagenet,lin2014microsoft,grauman2022ego4d}, datasets for VR problems in similar size remain relatively scarce to non-existent. Compounding this challenge is the complexity introduced by diverse hardware and input devices within and between manufacturers, which makes data collection and standardization even more intricate. Despite these obstacles, constructing comprehensive VR datasets holds immense potential. Such datasets could unlock progress in different areas. Examples of these areas are movement prediction \cite{bremer2021predicting,stein2022eye}, human motion tracking \cite{gamage2021so,welch2009history}, task prediction \cite{david2021towards,hu2020gaze}, gaze forecasting \cite{hu2021fixationnet,rolff2022gazetransformer}, locomotion \cite{gao2022eye}, or the development of systems for predicting motion sickness \cite{lee2019motion}. Furthermore, the availability of such extensive VR datasets would empower researchers to develop novel and innovative algorithms and gain more profound insights into behavioral patterns of humans. \\

Yet, a significant challenge in creating comprehensive VR datasets lies in developing a robust capturing system capable of handling the wide variety of head-mounted displays (HMDs) (e.g., Meta Quest or Apple Vision Pro), game engines (e.g., Unity3D or Unreal), and data modalities (e.g., IMU data from controller inputs, head position, gaze data, videos, depth data, and many more). 
On one side, such a system requires a robust data format and capturing tool that allows capturing a wide range of modalities from virtual environments (VEs) and hardware. On the other side, it should be as easily accessible as possible, enabling researchers from multiple disciplines to provide their data to the growing corpus of datasets. Fortunately, many modern HMDs provide a standardized OpenXR \cite{openxr} interface, exposing the functionality of each device. It was designed such, that it standardizes across multiple devices, including HMDs, controllers, base stations haptic devices, and other trackers, such as body trackers, eye trackers, or hand trackers. This upside of the converging OpenXR standard enables a data-capturing tool that can support any device supported by the standard itself.\\

In this paper, we introduce a toolkit for collecting such multimodal VR datasets using the Unity3D\footnote{\url{https://unity.com/}} game engine. We designed our toolkit such that it addresses the aforementioned challenges, seamlessly supporting any HMD compatible with OpenXR, and enabling the capture of most devices and features accessible through this standard. Furthermore, we provide an open data format used to store the captured data in an efficient and extensible way that can be externally analyzed through our provided Python toolkit. To summarize, we provide the following contributions:

\begin{itemize}
    \item We propose a frame-independent toolkit for the Unity 3D game engine, alongside a set of Python scripts, for the collection and analysis of vast datasets that can be used to train machine learning algorithms.
    \item We provide a data format, which allows encoding any multimodal data captured from any HMD supporting OpenXR.
    \item We provide guidelines for data collection to ensure correct, legal, and ethical data collection.
\end{itemize}

\section{Related Work}
Data collection within VEs is a fundamental element of most VR experiments \cite{lazar2017research,kjeldskov2003review}, serving as a critical instrument to validate or challenge the hypotheses of papers. For studies, involving human participants, it is not unusual to frequently employ standardized or custom-designed questionnaires, 
obtaining insights either during or after the experiment. The choice of questionnaire is typically tailored to the research domain and the specific hypothesis. In addition to questionnaires, interviews and other qualitative methods are often employed to provide a more profound understanding of human interaction with technology.\\

Another quite common research element is the recording of metrics during experiments, capturing both discrete and sequential data. This includes single-value metrics, for example, game scores \cite{karaosmanoglu2022canoe} or task completion speed \cite{kangas2022trade}, but also temporal data, like heart rate \cite{rockstroh2019virtual}, or eye movement data \cite{rolff2022gazetransformer}. 
Most likely, most studies performed will capture multiple modalities, including quantitative and qualitative data through questionnaires, metrics, videos, and other modalities \cite{hu2020gaze,hu2021fixationnet}.
These metrics may either be used on their own for the confirmation of a hypothesis or in combination with other qualitative measures. Together, these diverse data collection methods provide a robust foundation for understanding and interpreting human behavior in virtual settings.\\

However, some types of data acquisition might be more challenging than others or have to follow so determined process. To simplify this and to avoid errors, previous literature focuses on providing toolkits for the correct execution of the experimental procedure \cite{zenner2023staircasetoolkit,villenave2022xrecho,javerliat2024plume,cognitive3D}. For example, Zenner et al. \cite{zenner2023staircasetoolkit} provide a toolkit for the staircase procedure for estimating psychophysical detection thresholds. Other, toolkits, such as the \emph{Toggle toolkit} \cite{ugwitz2021toggle}, focus on providing a way to trigger specific events and actions inside the VE; however, their toolkit lacks full recording capabilities. In contrast, some proprietary tools such as \emph{NVIDIA Virtual Reality Capture and Replay} (\emph{NVIDIA VCR}) \cite{nvidiaVCR}, \emph{Tobii's Ocumen} \cite{tobiiOcumen}, \emph{cognitive3D} \cite{cognitive3D}, or open access tools, like \emph{XREcho}, aim at recording the full application state. Yet, these are often restricted to PC-VR or Web-VR, making them impractical to record for standalone VR (i.e., for devices such as the Meta Quest). To solve this gap, Javerliat et al. \cite{javerliat2024plume} present a more universal tool to record user behavior by capturing the state of the application for a wide range of devices. Their \emph{PLUME} toolkit~\cite{javerliat2024plume} provides an effortless way to capture a full experimental setup, with a wide variety of captured data.

\section{Toolkit}
In this section, we will introduce our \emph{\fulltoolkit (\toolkit)} toolkit. We base our toolkit on the Unity3D game engine.
Contrary to previous toolkits like XREcho \cite{villenave2022xrecho} or PLUME \cite{javerliat2024plume}, we do not want to capture the full state, but rather directly record the output of the hardware that is provided to the Unity game engine. This has two reasons: First, it allows us to capture the data at a fixed update rate, independent of the frame rate. This is especially important for captured data with high change rates or fast movements, such as eye movement or controller input data. Second, the data captured can be directly used for inference on a trained machine learning model, since the captured data can be directly used to train the model and does not deviate between capture and runtime.\\

\noindent For clarity, we split our toolkit into three different parts:
\begin{enumerate}
    \item A data format designed to capture any data provided by OpenXR, which also allows being easily extended for special devices that do not provide OpenXR support yet.
    \item A recording tool for the Unity3D game engine with a minimal setup that directly stores the OpenXR data provided to the Unity3D engine in our aforementioned data format.
    \item A set of Python scripts, that act as a toolset to generate machine learning data for external analysis.
\end{enumerate}

In the following, we will first discuss our data format. Afterward, we then go into the details of our data recording tool and give a short overview of the Python toolkit.

\subsection{Data Format}
\label{sec: Data Format}
\begin{figure}[t]
    \centering
    \includegraphics[width=1.0\columnwidth,trim={0 0 2.5cm 0},clip]{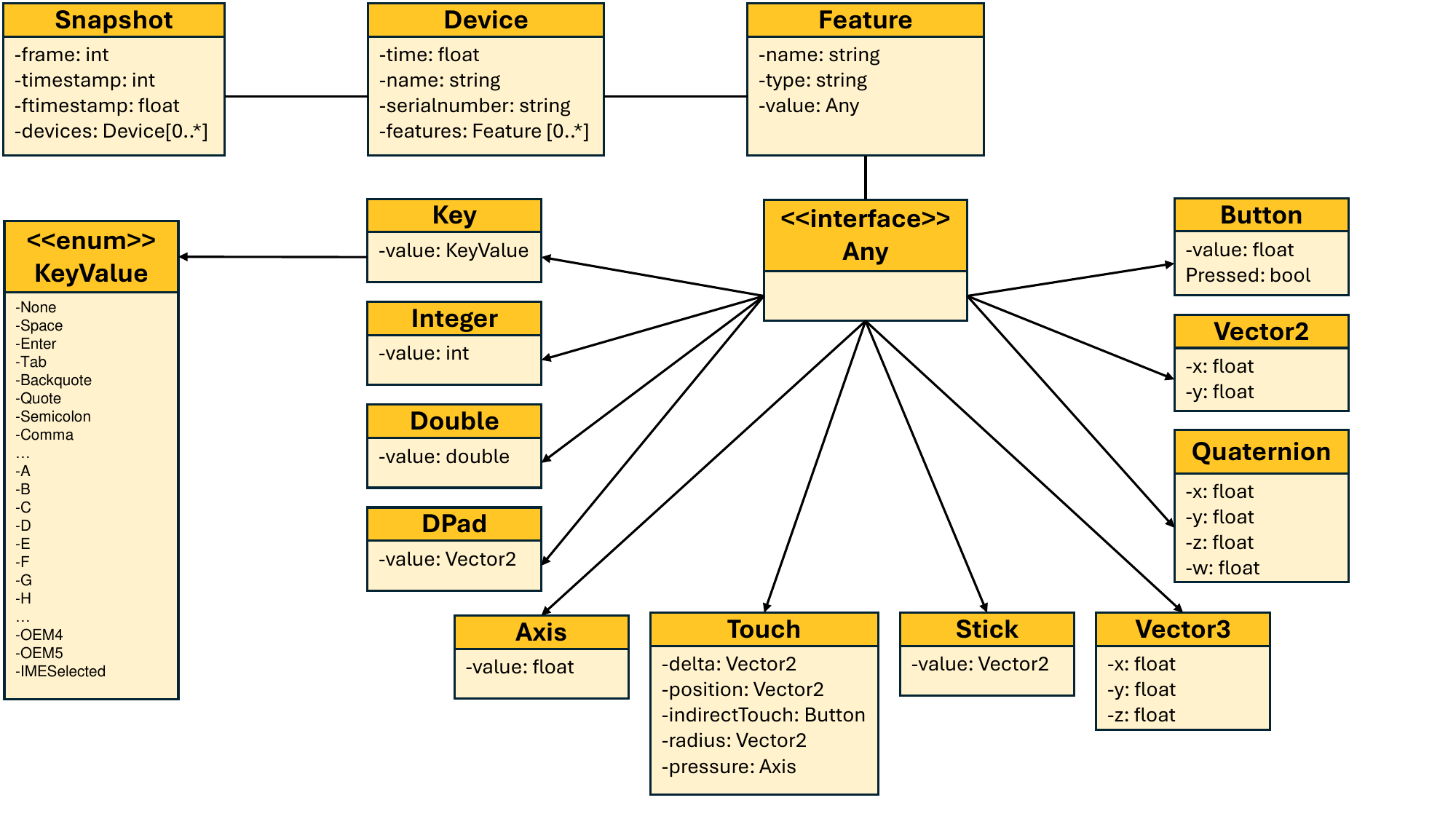}
    \caption{Data format used to store the captured data of our recording toolkit. See \cref{sec: Data Format} for more details.}
    \label{fig: UML}
\end{figure}
Before, detailing our capturing system, we want to provide an overview of our data format used to store any arbitrary data outputted by the Unity3D interface of OpenXR. Here, we propose two different types of data storage, providing a trade-off between data sizes and data handling. To store the data, we either provide Newline Delimited JSON (NDJSON)\footnote{\url{https://github.com/ndjson/ndjson-spec}} or binary storage through MessagePack\footnote{\url{https://msgpack.org/}}, which provides an efficient binary serialization format similar to the JSON format. To store all possible data types, we define a hierarchy of different structures (c.f.~\cref{fig: UML}):\\
\begin{figure*}[t]
    \centering
    \includegraphics[width=0.95\textwidth,trim={0 12.05cm 1.1cm 0},clip]{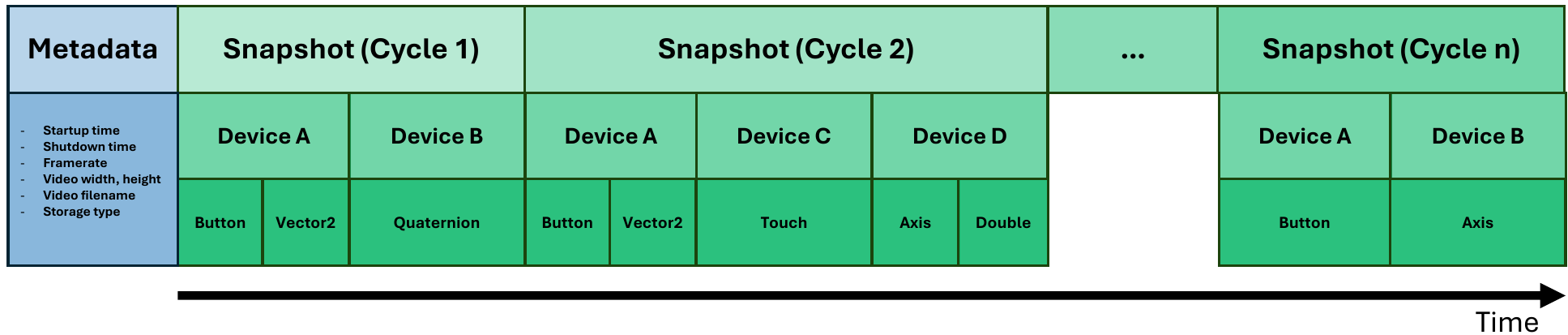}
    \caption{Example showing the data format. The first entry is always a metadata structure (c.f.~\cref{sec: Metadata}) followed by multiple snapshots (c.f.~\cref{sec: Snapshot} containing an array of devices. Note that it is not necessary for a snapshot to contain the same devices every update cycle. Furthermore, each device contains a set of features (c.f.~\cref{sec: controls}) that are also not restricted to the same layout every update cycle.}
    \label{fig: data format}
\end{figure*}
\subsubsection{Metadata}
\label{sec: Metadata}
The initial entry in our data format is defined by a series of metadata elements (see \cref{fig: data format}) that provide essential information for external analysis and validation. These metadata entries include details such as the startup time, end time, and framerate/polling rate, enabling the alignment of the recorded data. Additionally, attributes such as video resolution (width and height) and the corresponding filename are stored. To ensure the validity of the recordings, we also incorporate details about the utilized HMD and the storage medium.

\subsubsection{Snapshot}
\label{sec: Snapshot}
Our tool is specifically designed to integrate seamlessly with Unity3D; hence aligning its data structures with the Unity3D Input System \cite{inputSystem}. At the core, we use a snapshot structure, representing all data captured during a single update cycle. We define an update cycle as the interval in which the input system updates according to a specified polling rate. Each snapshot encapsulates information about the  associated frame, timestamp, and device metadata. Since multiple devices can be updated within the same cycle, a single snapshot may represent data from multiple devices. This structured approach ensures that our data format efficiently stores precisely one snapshot for every input system update cycle.

\subsubsection{Device}
\label{sec: Device}
A device refers to a physical or virtual hardware component being recorded, whether partially or in its entirety. This might be, for example, a controller, eye tracker, or the HMD. A device provides metadata such as timestamps, names, and serial numbers to enable identification. Additionally, each device is associated with a set of features that encapsulate and represent its hardware capabilities. To accommodate diverse use cases, we propose a flexible data abstraction layer designed to (i) integrate seamlessly with the Unity Input System (refer to \cref{sec: controls}) and (ii) support the extension of custom user-defined virtual device types, even for those that do not natively expose an OpenXR interface to Unity.

\subsubsection{Feature and value types}
\label{sec: controls}
A feature represents an abstraction of a Unity3D Input System control type\footnote{\url{https://docs.unity3d.com/Packages/com.unity.inputsystem@1.0/manual/Controls.html}}, encompassing its name, type, and associated data. As previously mentioned, to support extensibility and flexibility, our approach provides an abstraction layer for the value data that stores the captured data, allowing us to integrate custom user-defined types while maintaining compatibility with the control types offered by the Unity3D Input System; hence requiring us to store the type of the data for reconstruction during external analysis.\\

By default, the system supports fundamental data types, such as Integer and Double, ensuring the representation and storage of numerical information. Beyond these, we include commonly used mathematical constructs like Vector2, Vector3, and Quaternions. All these already allow for storing complex input systems, such as controllers, eye trackers, or heart rate sensors. For ease of use, we also add the missing input options of the Unity3D Input System, supporting controls such as Axis, Button, Key, Stick, DPad, and Touch, thereby offering a wide range of input scenarios.\\

Given all outlined value types, it enables us to define physical or virtual hardware as a composition of multiple control types. 
A basic controller, equipped with three buttons, a touchpad, and two triggers, can be characterized by five button controls (corresponding to the three primary buttons and two trigger buttons), a touch control for the touchpad, a Vector3 control representing its positional coordinates, and a Quaternion control capturing its rotational orientation.

\subsection{Data Collection Toolkit \& Procedure}
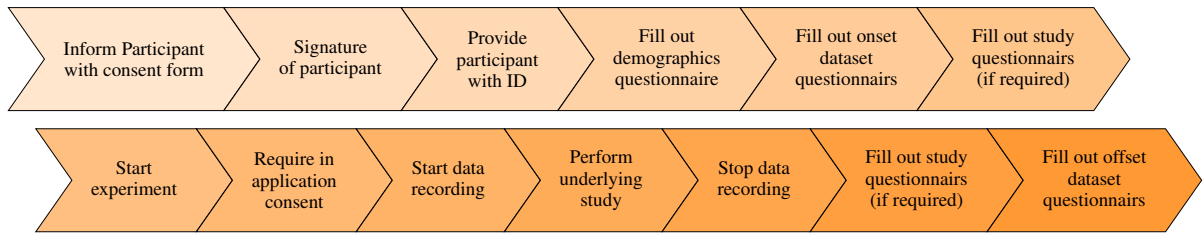
\begin{figure*}[t]
    \centering
    \resizebox{0.9\textwidth}{!}{
\begin{tikzpicture}
  \tikzset{
    bound/.style={
      draw,
      minimum height=2cm,
      inner sep=1em,
    },
    arrow/.style={
      draw,
      minimum height=1.7cm,
      inner sep=1em,
      shape=signal,
      signal from=west,
      signal to=east,
      signal pointer angle=110,
    }
  }
  \begin{scope}[start chain=transition going right,node distance=-\pgflinewidth]
    \node[arrow,on chain,align=center, fill=orange!20] {Inform Participant\\with consent form};
    \node[arrow,on chain,align=center, fill=orange!25] {Signature\\of participant};
    \node[arrow,on chain,align=center, fill=orange!30] {Provide\\participant\\with ID};
    \node[arrow,on chain,align=center, fill=orange!35] {Fill out\\demographics\\ questionnaire};
    \node[arrow,on chain,align=center, fill=orange!40] {Fill out onset\\dataset\\questionnairs};
    \node[arrow,on chain,align=center, fill=orange!45] {Fill out study\\questionnairs\\(if required)};
  \end{scope}

  \begin{scope}[yshift=-2cm,start chain=transition going right,node distance=-\pgflinewidth]
    \node[arrow,on chain,align=center, fill=orange!50] {Start\\experiment};
    \node[arrow,on chain,align=center, fill=orange!55] {Require in\\application\\ consent};
    \node[arrow,on chain,align=center, fill=orange!60] {Start data\\recording};
    \node[arrow,on chain,align=center, fill=orange!65] {Perform\\underlying\\study};
    \node[arrow,on chain,align=center, fill=orange!70] {Stop data\\recording};
    \node[arrow,on chain,align=center, fill=orange!75] {Fill out study\\questionnairs\\(if required)};
    \node[arrow,on chain,align=center, fill=orange!80] {Fill out offset\\dataset\\questionnairs};
  \end{scope}
\end{tikzpicture}
    }
    \caption{Predefined study procedure when capturing data through our toolkit.}
    \label{fig: Collection Procedure}
\end{figure*}
In this section, we detail a standardized approach for data collection using our toolkit alongside its technical details. While this is not the only possibility to collect data, we want to provide a guideline to ensure a standardized, ethical, and correctly captured dataset. For a detailed overview, see~\cref{fig: Collection Procedure}. 

\subsubsection{Ethical and Data Protection Considerations:}

We designed our data collection system from the ground up to be as ethical as possible, aligning with established recommendations 
\cite{richards2014big, boppiniti2023data}. Therefore, we want to outline some guidelines for the ethical use of our system.
Before initiating the recording, all participants must be thoroughly informed about the nature and purpose of the data being collected via a detailed consent form. The consent form must educate participants about the potential authorized users that will have access to the data. Participants are required to provide written consent, which is tied to a unique identifier to facilitate their rights to data deletion, in full compliance with the General Data Protection Regulation (GDPR)\footnote{\url{https://eur-lex.europa.eu/eli/reg/2016/679/oj}} enacted by the European Union. To reinforce the transparency and autonomy of participants and the developers of the captured application, a second consent confirmation step should be performed within the captured application, following the recommendation of our institutional ethics committee. This step requires participants to actively affirm their agreement via a button press. All raw data collected is securely stored on a GDPR-compliant server located within the European Union, ensuring rigorous adherence to data protection standards.

\subsubsection{Standardized Surveys:}
In addition to the collection of quantitative data, we incorporate a series of questionnaires into our toolkit to establish connections between various data modalities and quantitative metrics. This approach enables multimodal data analysis without requiring a purpose-built experimental setup, while also providing opportunities to generate novel hypotheses from pre-existing datasets. For example, this would allow replicating already performed studies, such as the finding that mental workload can be associated with pupil diameter \cite{walter2021cognitive}. Given the wide range of available surveys, we selected questionnaires that are broadly applicable across diverse VR studies, ensuring their applicability for data collection:\\

To ensure consistency in data collection, we begin with a standardized demographics' questionnaire (see \cref{sec: Demographics}). This gathers essential details such as participants' age, gender, native language, use of vision correction, and prior experience with VR. Additionally, we recommend administering the NASA-TLX \cite{hart2006nasa}, the Simulator Sickness Questionnaire (SSQ) \cite{kennedy1993simulator}, and the Igroup Presence Questionnaire (IPQ) \cite{schubert2001experience} both at the study’s beginning (onset) and ending (offset). These tools may provide valuable insights into participants' experiences and ensure a comprehensive assessment of qualitative metrics without an extensive time impact. 

\subsubsection{Technical Details:}
In contrast to many data capture toolkits, we designed our system such that it runs independent of the frame rate of the application. This is critical, as a stutter due to a longer performed rendering computation of the graphics card may have an impact on the captured data. This may influence the predictive performance of any deep-learning (DL) model. 
As we want to support using the captured output directly for training and inference, frame dependence may make a difference, for example, due to changed computation or different scenes, resulting in differently aligned data. Furthermore, frame dependence also adds another layer of complexity that needs to be taken care of as it may require performing multiple alignments due to multiple reported data points for a singular frame (i.e., gaze data from an eye tracker).\\

\begin{figure}[t]
    \centering
    \resizebox{0.9\columnwidth}{!}{
\begin{tikzpicture}
  \tikzset{
    bound/.style={
      draw,
      minimum height=2cm,
      inner sep=1em,
    },
    arrow/.style={
      draw,
      minimum height=1.7cm,
      inner sep=1em,
      shape=signal,
      signal from=west,
      signal to=east,
      signal pointer angle=110,
    }
  }
  \begin{scope}[start chain=transition going right,node distance=-\pgflinewidth]
    \node[arrow,on chain,align=center, fill=orange!20] {Integrate\\toolkit into\\ application};
    \node[arrow,on chain,align=center, fill=orange!25] {Perform\\data\\collection};
    \node[arrow,on chain,align=center, fill=orange!30] {Store raw data};
  \end{scope}
  
  \begin{scope}[yshift=-2cm,start chain=transition going right,node distance=-\pgflinewidth]
    \node[arrow,on chain,align=center, fill=orange!35] {Filter data\\through\\python};
    \node[arrow,on chain,align=center, fill=orange!40] {Analyze data\\or\\train model};
    \node[arrow,on chain,align=center, fill=orange!45] {Confirm/deny \\hypothesis};
  \end{scope}
\end{tikzpicture}
    }
    \caption{Overall procedure for our toolkit.}
    \label{fig: General Procedure}
\end{figure}
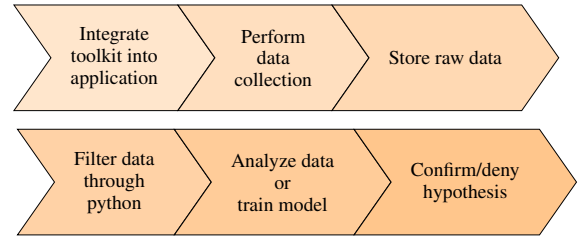

Beyond that, we do not want users of the collection system to specify the intricate details of the captured details and captured objects. Instead, we fully collect all connected hardware, even including the keyboard and mouse in a PC-VR setup. Implementation-wise, we hook directly into the input system and specify a polling frequency at which new device information, will be reported. As a result, we capture all events sent to Unity, including OpenXR input events. This allows us to filter for necessary hardware later through Python (c.f.~\cref{fig: General Procedure}).\\

Due to some hardware not supporting all OpenXR features, we provide implementations for virtual devices, such as an eye tracker for the Meta Quest Pro, Pico Neo3 Pro, or the HTC Vive Pro Eye. We also support capturing video on a subset of hardware. 

\subsubsection{Analysis Pipeline}
Our analysis pipeline is based on python and implements all data types supported by our capturing system (c.f.~\cref{sec: Data Format}). This allows users to easily parse files outputted by the system and their conversion to different file formats, such as comma separated values~(CSV) for data analysis or model training.

\section{Discussion \& Limitations}
This paper introduces a versatile data collection toolkit designed specifically for virtual reality studies. The proposed toolkit has a streamlined integration, offers frame rate-independent recording, and provides a standardized methodology for data acquisition. Central to this approach is the inclusion of ethical guidelines and standardized questionnaires to ensure robust and reproducible research practices. Furthermore, we provide a Python-based analysis tool for processing the defined data format. By incorporating frame rate independence, our framework eliminates the need for distinct implementations when capturing or interfacing with deep learning models, enhancing both efficiency and scalability.\\

Despite its strengths, our work has certain limitations worth noting. Foremost among them is that the current iteration of our toolbox is restricted to supporting Unity. Expanding this compatibility could be achieved by implementing an interceptor plugin for OpenXR, which would enable the interception of all native OpenXR calls\footnote{\url{https://registry.khronos.org/OpenXR/specs/1.0/loader.html}}. Additionally, while our framework excels in its targeted simplicity, it does not capture the complete state of the application, unlike solutions such as PLUME \cite{javerliat2024plume}. This design trade-off means that, at present, we cannot automatically capture in-application metrics—such as game scores -- without integrating custom virtual devices.


\acknowledgments{
The research for this paper was funded by the Deutsche Forschungsgemeinschaft (DFG, German Research Foundation) -- project no. 511498220 at the University of Hamburg.
}

\clearpage

\bibliographystyle{abbrv-doi}
\bibliography{template}

\begin{thebibliography}{10}

\bibitem{cognitive3D}
{cognitive3d}.
\newblock \url{https://cognitive3d.com/}.
\newblock [Online; accessed 30-November-2024].

\bibitem{nvidiaVCR}
{NVIDIA Virtual Reality Capture and Replay}.
\newblock \url{https://docs.nvidia.com/vcr-sdk/overview/overview.html/}.
\newblock [Online; accessed 30-November-2024].

\bibitem{tobiiOcumen}
{Tobii Ocumen}.
\newblock \url{https://developer.tobii.com/xr/solutions/tobii-ocumen/}.
\newblock [Online; accessed 30-November-2024].

\bibitem{inputSystem}
{Unity Input System}.
\newblock \url{https://docs.unity3d.com/Packages/com.unity.inputsystem@1.0}.
\newblock [Online; accessed 30-November-2024].

\bibitem{openxr}
{OpenXR}.
\newblock \url{https://www.khronos.org/api/index_2017/openxr}, 2017.
\newblock [Online; accessed 28-November-2024].

\bibitem{boppiniti2023data}
S.~T. Boppiniti.
\newblock Data ethics in ai: Addressing challenges in machine learning and data governance for responsible data science.
\newblock {\em International Scientific Journal for Research}, 5(5), 2023.

\bibitem{bremer2021predicting}
G.~Bremer, N.~Stein, and M.~Lappe.
\newblock Predicting future position from natural walking and eye movements with machine learning.
\newblock In {\em 2021 IEEE international conference on artificial intelligence and virtual reality (AIVR)}, pp. 19--28. IEEE, 2021.

\bibitem{david2021towards}
B.~David-John, C.~Peacock, T.~Zhang, T.~S. Murdison, H.~Benko, and T.~R. Jonker.
\newblock Towards gaze-based prediction of the intent to interact in virtual reality.
\newblock In {\em ACM symposium on eye tracking research and applications}, pp. 1--7, 2021.

\bibitem{gamage2021so}
N.~M. Gamage, D.~Ishtaweera, M.~Weigel, and A.~Withana.
\newblock So predictable! continuous 3d hand trajectory prediction in virtual reality.
\newblock In {\em The 34th Annual ACM Symposium on User Interface Software and Technology}, pp. 332--343, 2021.

\bibitem{gao2022eye}
H.~Gao and E.~Kasneci.
\newblock Eye-tracking-based prediction of user experience in vr locomotion using machine learning.
\newblock In {\em Computer Graphics Forum}, vol.~41, pp. 589--599. Wiley Online Library, 2022.

\bibitem{grauman2022ego4d}
K.~Grauman, A.~Westbury, E.~Byrne, Z.~Chavis, A.~Furnari, R.~Girdhar, J.~Hamburger, H.~Jiang, M.~Liu, X.~Liu, et~al.
\newblock Ego4d: Around the world in 3,000 hours of egocentric video.
\newblock In {\em Proceedings of the IEEE/CVF Conference on Computer Vision and Pattern Recognition}, pp. 18995--19012, 2022.

\bibitem{hart2006nasa}
S.~G. Hart.
\newblock Nasa-task load index (nasa-tlx); 20 years later.
\newblock In {\em Proceedings of the human factors and ergonomics society annual meeting}, vol.~50, pp. 904--908. Sage publications Sage CA: Los Angeles, CA, 2006.

\bibitem{hu2020gaze}
Z.~Hu.
\newblock Gaze analysis and prediction in virtual reality.
\newblock In {\em 2020 IEEE conference on virtual reality and 3D user interfaces abstracts and workshops (VRW)}, pp. 543--544. IEEE, 2020.

\bibitem{hu2021fixationnet}
Z.~Hu, A.~Bulling, S.~Li, and G.~Wang.
\newblock Fixationnet: Forecasting eye fixations in task-oriented virtual environments.
\newblock {\em IEEE Transactions on Visualization and Computer Graphics}, 27(5):2681--2690, 2021.

\bibitem{javerliat2024plume}
C.~Javerliat, S.~Villenave, P.~Raimbaud, and G.~Lavou{\'e}.
\newblock Plume: Record, replay, analyze and share user behavior in 6dof xr experiences.
\newblock {\em IEEE Transactions on Visualization and Computer Graphics}, 2024.

\bibitem{kangas2022trade}
J.~Kangas, S.~K. Kumar, H.~Mehtonen, J.~J{\"a}rnstedt, and R.~Raisamo.
\newblock Trade-off between task accuracy, task completion time and naturalness for direct object manipulation in virtual reality.
\newblock {\em Multimodal Technologies and Interaction}, 6(1):6, 2022.

\bibitem{karaosmanoglu2022canoe}
S.~Karaosmanoglu, L.~Kruse, S.~Rings, and F.~Steinicke.
\newblock Canoe vr: An immersive exergame to support cognitive and physical exercises of older adults.
\newblock In {\em CHI Conference on Human Factors in Computing Systems Extended Abstracts}, pp. 1--7, 2022.

\bibitem{kennedy1993simulator}
R.~S. Kennedy, N.~E. Lane, K.~S. Berbaum, and M.~G. Lilienthal.
\newblock Simulator sickness questionnaire: An enhanced method for quantifying simulator sickness.
\newblock {\em The international journal of aviation psychology}, 3(3):203--220, 1993.

\bibitem{kjeldskov2003review}
J.~Kjeldskov and C.~Graham.
\newblock A review of mobile hci research methods.
\newblock In {\em International Conference on Mobile Human-Computer Interaction}, pp. 317--335. Springer, 2003.

\bibitem{lazar2017research}
J.~Lazar, J.~H. Feng, and H.~Hochheiser.
\newblock {\em Research methods in human-computer interaction}.
\newblock Morgan Kaufmann, 2017.

\bibitem{lee2019motion}
T.~M. Lee, J.-C. Yoon, and I.-K. Lee.
\newblock Motion sickness prediction in stereoscopic videos using 3d convolutional neural networks.
\newblock {\em IEEE transactions on visualization and computer graphics}, 25(5):1919--1927, 2019.

\bibitem{lin2014microsoft}
T.-Y. Lin, M.~Maire, S.~Belongie, J.~Hays, P.~Perona, D.~Ramanan, P.~Doll{\'a}r, and C.~L. Zitnick.
\newblock Microsoft coco: Common objects in context.
\newblock In {\em Computer Vision--ECCV 2014: 13th European Conference, Zurich, Switzerland, September 6-12, 2014, Proceedings, Part V 13}, pp. 740--755. Springer, 2014.

\bibitem{patel2020introduction}
J.~M. Patel and J.~M. Patel.
\newblock Introduction to common crawl datasets.
\newblock {\em Getting structured data from the internet: running web crawlers/scrapers on a big data production scale}, pp. 277--324, 2020.

\bibitem{raffel2020exploring}
C.~Raffel, N.~Shazeer, A.~Roberts, K.~Lee, S.~Narang, M.~Matena, Y.~Zhou, W.~Li, and P.~J. Liu.
\newblock Exploring the limits of transfer learning with a unified text-to-text transformer.
\newblock {\em Journal of machine learning research}, 21(140):1--67, 2020.

\bibitem{richards2014big}
N.~M. Richards and J.~H. King.
\newblock Big data ethics.
\newblock {\em Wake Forest L. Rev.}, 49:393, 2014.

\bibitem{rockstroh2019virtual}
C.~Rockstroh, J.~Blum, and A.~S. G{\"o}ritz.
\newblock Virtual reality in the application of heart rate variability biofeedback.
\newblock {\em International Journal of Human-Computer Studies}, 130:209--220, 2019.

\bibitem{rolff2022gazetransformer}
T.~Rolff, H.~M. Harms, F.~Steinicke, and S.~Frintrop.
\newblock Gazetransformer: Gaze forecasting for virtual reality using transformer networks.
\newblock In {\em DAGM German Conference on Pattern Recognition}, pp. 577--593. Springer, 2022.

\bibitem{russakovsky2015imagenet}
O.~Russakovsky, J.~Deng, H.~Su, J.~Krause, S.~Satheesh, S.~Ma, Z.~Huang, A.~Karpathy, A.~Khosla, M.~Bernstein, et~al.
\newblock Imagenet large scale visual recognition challenge.
\newblock {\em International journal of computer vision}, 115:211--252, 2015.

\bibitem{schubert2001experience}
T.~Schubert, F.~Friedmann, and H.~Regenbrecht.
\newblock The experience of presence: Factor analytic insights.
\newblock {\em Presence: Teleoperators \& Virtual Environments}, 10(3):266--281, 2001.

\bibitem{stein2022eye}
N.~Stein, G.~Bremer, and M.~Lappe.
\newblock Eye tracking-based lstm for locomotion prediction in vr.
\newblock In {\em 2022 IEEE conference on virtual reality and 3D user interfaces (VR)}, pp. 493--503. IEEE, 2022.

\bibitem{ugwitz2021toggle}
P.~Ugwitz, A.~{\v{S}}a{\v{s}}inkov{\'a}, {\v{C}}.~{\v{S}}a{\v{s}}inka, Z.~Stacho{\v{n}}, and V.~Ju{\v{r}}{\'\i}k.
\newblock Toggle toolkit: A tool for conducting experiments in unity virtual environments.
\newblock {\em Behavior research methods}, pp. 1--11, 2021.

\bibitem{villenave2022xrecho}
S.~Villenave, J.~Cabezas, P.~Baert, F.~Dupont, and G.~Lavou{\'e}.
\newblock Xrecho: A unity plug-in to record and visualize user behavior during xr sessions.
\newblock In {\em Proceedings of the 13th ACM Multimedia Systems Conference}, pp. 341--346, 2022.

\bibitem{walter2021cognitive}
K.~Walter and P.~Bex.
\newblock Cognitive load influences oculomotor behavior in natural scenes.
\newblock {\em Scientific Reports}, 11(1):12405, Jun 2021. doi: {{%
10\hspace{.1pt}\discretionary{.}{%
}{.}\hspace{.4pt}1038\discretionary{/}{%
}{/}s41598\discretionary{%
}{-}{-}021\discretionary{%
}{-}{-}91845\discretionary{%
}{-}{-}5}}


\bibitem{weber2024redpajama}
M.~Weber, D.~Fu, Q.~Anthony, Y.~Oren, S.~Adams, A.~Alexandrov, X.~Lyu, H.~Nguyen, X.~Yao, V.~Adams, et~al.
\newblock Redpajama: an open dataset for training large language models.
\newblock {\em arXiv preprint arXiv:2411.12372}, 2024.

\bibitem{welch2009history}
G.~F. Welch.
\newblock History: The use of the kalman filter for human motion tracking in virtual reality.
\newblock {\em Presence}, 18(1):72--91, 2009.

\bibitem{zenner2023staircasetoolkit}
A.~Zenner, K.~Ullmann, C.~Karr, O.~Ariza, and A.~Kr\"{u}ger.
\newblock The staircase procedure toolkit: Psychophysical detection threshold experiments made easy.
\newblock In {\em Proceedings of the 29th ACM Symposium on Virtual Reality Software and Technology}, VRST '23. Association for Computing Machinery, New York, NY, USA, 2023. doi: {{%
10\hspace{.1pt}\discretionary{.}{%
}{.}\hspace{.4pt}1145\discretionary{/}{%
}{/}3611659\hspace{.1pt}\discretionary{.}{%
}{.}\hspace{.4pt}3617218}}


\end{thebibliography}

\clearpage
\appendix
\appendixpage
\addappheadtotoc
\onecolumn
\section{Demographics Questionnaire}
\label{sec: Demographics}
\begin{table}[!h]
    \centering
    \resizebox{\textwidth}{!}{%
    \begin{tabular}{p{6.5cm}p{8.9cm}}
        What is your age in years? (please fill in a number): & \\
        & \hrulefill
        \\\\\\\\
        What gender do you identify as? & \begin{itemize}[label=$\circ$]
            \item female
            \item male
            \item non-binary
            \item prefer to not disclose
            \item prefer to self-describe\hspace{0.5cm} \hrulefill
        \end{itemize}
        \\\\\\\\
        Please specify your native language. If you are multilingual, please specify your preferred language: & \\
        & \hrulefill
        \\\\\\\\
        Do you need vision correction? & \begin{itemize}[label=$\circ$]
            \item Yes
            \item No
        \end{itemize}
        \\\\\\\\
        How would you rate your experience in VR? & \\\\
        \hspace{0.5cm}
        \begin{tabular}{cccccccc}
              \small{no experience}
            & \small{less than once a year}
            & \small{once a year}
            & \small{every six months}
            & \small{every three months}
            & \small{once a month}
            & \small{once a week}
            & \small{daily VR user}\\
            $\circ$ & $\circ$ & $\circ$ & $\circ$ & $\circ$ & $\circ$ & $\circ$ & $\circ$
        \end{tabular}%
        \\\\\\\\
    \end{tabular}%
    }
    \caption{Our proposed standardized demographics questionnaire for data collection.}
    \label{tab: Demographics}
\end{table}

\newpage


\acknowledgments{
ChatGPT was used as a tool for rephrasing parts of this paper. As a prompt: ``\emph{Can you rewrite this \{section\} for a paper to sound more scientific, enjoyable and engaging without altering its meaning: \{text\}}'' was used, replacing \{section\} with the corresponding section title and \{text\} with the text to be rephrased.
}

\end{document}